\documentclass{article}

\usepackage{PRIMEarxiv}

\usepackage[utf8]{inputenc} 
\usepackage[T1]{fontenc}    
\usepackage{hyperref}       
\usepackage{url}            
\usepackage{booktabs}       
\usepackage{amsfonts}       
\usepackage{nicefrac}       
\usepackage{microtype}      
\usepackage{fancyhdr}       
\usepackage{graphicx}       
\graphicspath{{media/}}     

\RequirePackage{algorithm}
\RequirePackage{algorithmic}

\usepackage[numbers]{natbib}
\usepackage{amsmath}
\usepackage{amssymb}
\usepackage{mathtools}
\usepackage{amsthm}
\usepackage{optidef}
\usepackage{thmtools}
\usepackage{thm-restate}
\usepackage[capitalize,noabbrev]{cleveref}
\usepackage{wrapfig}

\usepackage{xcolor}         
\usepackage{graphicx}   
\usepackage{wrapfig}
\usepackage{subcaption}
\usepackage{enumitem}
\usepackage{amsmath}
\newtheorem{definition}{Definition}

\usepackage{multirow}
\usepackage{cleveref}
\usepackage{tabularx}

\title{Dual Node and Edge Fairness-Aware Graph Partition
}

\author{
  \vspace{1mm} Tingwei Liu, Peizhao Li, and Hongfu Liu \\
  \vspace{1mm} Brandeis University \\
  \texttt{\{tingweiliu,peizhaoli,hongfuliu\}@brandeis.edu} \\
}

\begin{document}
\maketitle

\begin{abstract}
Fair graph partition of social networks is a crucial step toward ensuring fair and non-discriminatory treatments in unsupervised user analysis. Current fair partition methods typically consider \textit{node balance}, a notion pursuing a proportionally balanced number of nodes from all demographic groups, but ignore the bias induced by imbalanced edges in each cluster. To address this gap, we propose a notion \textit{edge balance} to measure the proportion of edges connecting different demographic groups in clusters. We analyze the relations between \textit{node balance} and \textit{edge balance}, then with line graph transformations, we propose a co-embedding framework to learn dual node and edge fairness-aware representations for graph partition. We validate our framework through several social network datasets and observe balanced partition in terms of both nodes and edges along with good utility. Moreover, we demonstrate our fair partition can be used as pseudo labels to facilitate graph neural networks to behave fairly in node classification and link prediction tasks.
\end{abstract}



\section{Introduction}
\label{sec:intro}

\begin{wrapfigure}[19]{r}{0.3\textwidth}\vspace{-5mm}
\includegraphics[width=1.0\linewidth]{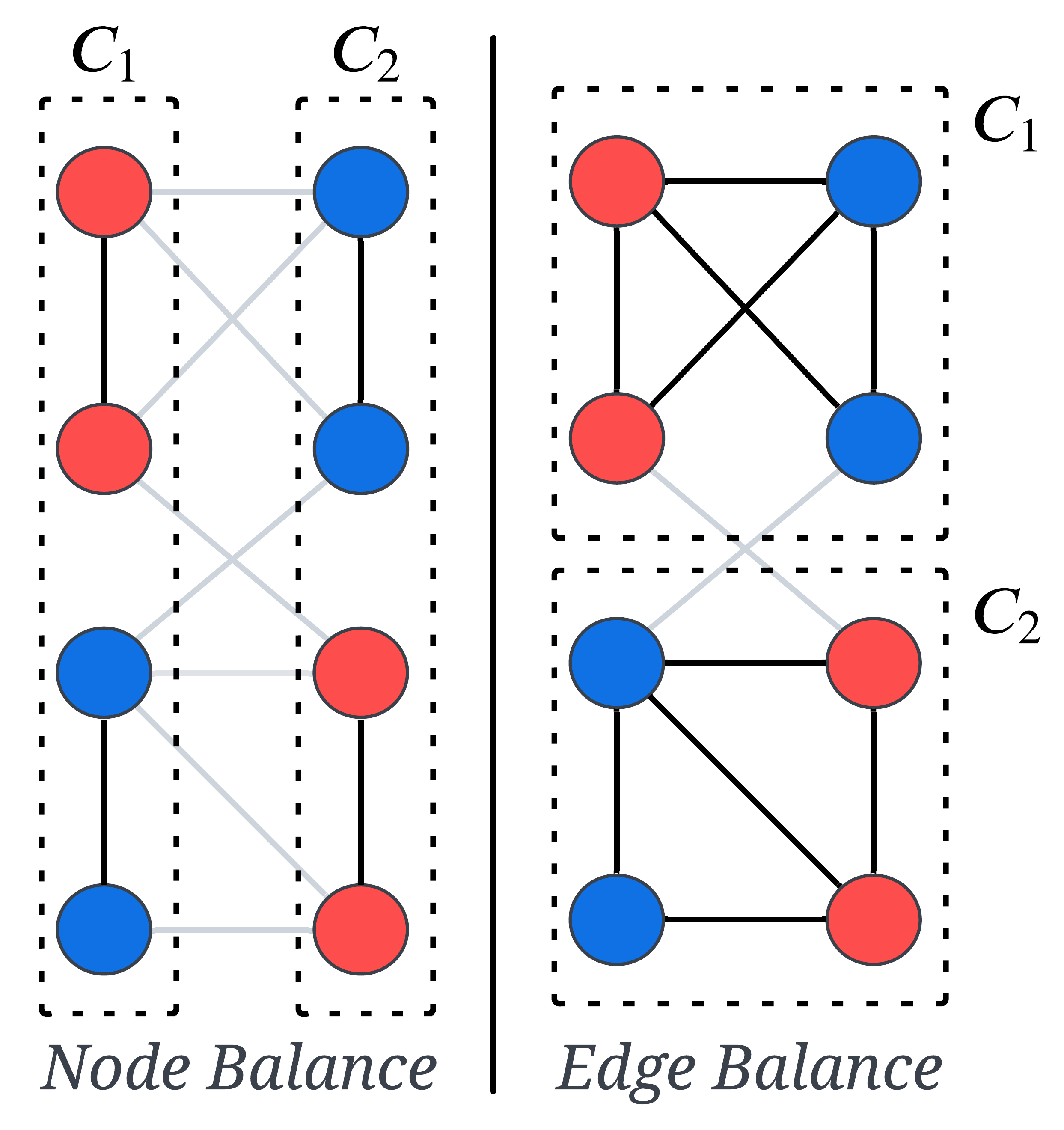}\vspace{-1mm}
\caption{Partitions showcasing node and edge balance constraints. $C_1$ and $C_2$ are two clusters in a graph partition. Colors in nodes indicate the demographic group.}
\label{fig:intro}
\end{wrapfigure}

Graph mining~\cite{nickel2015review,wu2020comprehensive,zhang2020deep} is the process to discover useful information and patterns from complex networks of interconnected data. Leveraging the innate and flexible nature of graphs to represent intricate relationships, this field has found applications in various domains, including social network analysis~\cite{sahoo2021multiple}, bioinformatics~\cite{zhang2021graph}, cybersecurity~\cite{hozhabrierdi2022coverd,chhabra2022robust}, and transportation systems~\cite{jiang2022graph}, among others. Graph partition is the technique used in graph mining to segment large graphs into disjoint clusters without any annotation, while nodes in each cluster enjoy similarity and remain densely connected, and nodes between partitions are sparsely connected. Graph partition can be used in unsupervised community detection to conduct analysis on users in social networks \cite{bedi2016community, lopes2020parallel}.

With the increasing reliance on graph mining techniques for human-related decision-making, concerns about fairness, bias, and discrimination in algorithmic outcomes have become progressively more relevant~\cite{dong2023fairness}. To meet the fairness criteria, fair graph partition~\cite{ahmadian2020fair,backurs2019scalable,bera2019fair,chierichetti2017fair,kleindessner2019guarantees} aims to conduct unbiased partition that keeps equal representations from all demographic groups in each cluster. As one of the realizations of fairness, current fair partition methods typically consider \textit{node balance}, i.e., each graph partition contains a proportionally balanced number of nodes from all demographic groups~\cite{chierichetti2017fair}. Though intuitive, we observe that only considering the \textit{node balance} constraint but ignoring the balance on edges is not enough to deliver sufficiently fair and meaningful partitions. We use~\Cref{fig:intro} to illustrate \textit{node balance} and the proposed notion \textit{edge balance}. Suppose the entire graph contains an equal number of nodes in red and blue demographic groups. The partitions $C_1$ and $C_2$ in the left figure contain an equal number of red and blue nodes such that a perfect \textit{node balance} is achieved. However, the two demographic groups in each partition are completely separated with no edge connection appearing between groups. With a closer look at each partition, the bias in edges can easily reveal the demographic information, therefore violating the intention of fairness. In the right figure, the two partitions are not only balanced in nodes, but also include balanced edges (2 edges within the same demographic group, and 3 across groups). The denser connections across demographic groups imply a meaningful partition with more frequent user interactions. 

\vspace{-3mm}
\paragraph{Contributions.} In this paper, driven by the limitations observed in current fair partition approaches, we raise the notion \textit{edge balance} to contribute to all-around fairness in unsupervised graph partition. We design \textit{edge balance} as a quantitative measure to assess the proportion of edges connecting different demographic groups and within the same demographic group in clusters. Upon this notion, we make the following significant contributions:
\setlist[itemize]{leftmargin=*}
\begin{itemize}
\vspace{-1mm}
    \item With graph generation models, we reveal the correlations between node balance and edge balance in graph partition and show how arbitrarily bad edge balance could be at certain levels of node balance. This observation indicates pursuing node balance solely is not sufficient for a fair graph partition. 
    \item With line graph transformations, we propose an innovative co-embedding framework designed to meticulously remove demographic information in both node and edge embedding using adversarial learning and let models effectively learn both good and dual fairness-aware representations for graph partition.
    \item  We demonstrate the remarkable efficacy of our proposed model by conducting comprehensive, methodical experiments on five wildly-used social network datasets. Moreover, our dual fair partitions effectively promote impartiality in critical downstream tasks, including node classification and link prediction.

\end{itemize}


\section{Related Work}

We introduce relevant works in fair clustering and fair graph mining, and highlight our major differences from the existing literature.

\vspace{-3mm}
\paragraph{Fairness with Data Partition.} Data partition, or called clustering, is a widely used unsupervised learning paradigm seeking to separate data into disjoint groups based on their similarity. With fairness considerations, fair clustering aims to ensure that the obtained clusters do not disproportionately represent demographic groups, such that each cluster is \textit{balanced}: the proportion of each group in a cluster adheres to the general demographic distribution in the entire dataset~\cite{chierichetti2017fair}. Algorithmic solutions have been proposed in clustering subject to fairness constraints. The pre-processing method uses Fairlets Decomposition~\cite{chierichetti2017fair} to partition the original dataset into smaller balanced subsets, then use \textit{k}-center or \textit{k}-means clustering to generate final fair clusters. However, the Fairlets Decomposition algorithm takes quadratic running time on the first sub-partition steps, therefore motivates subsequent research focus on limit time cost~\cite{backurs2019scalable, chakrabarty2021better} and bounded trade-off between fairness and group utilitarian~\cite{esmaeili2021fair}. Other works have also extended the learning paradigm beyond traditional clustering, such as deep clustering~\cite{li2020deep}, spectral clustering~\cite{kleindessner2019guarantees}, and hierarchical clustering~\cite{ahmadian2020fair}.

\vspace{-3mm}
\paragraph{Fair Graph Mining.} Research works in fair graph mining offer algorithmic insights into getting useful and non-discriminatory predictions in graph-structured data. Numerous studies have emerged in various directions~\cite{dong2023fairness,kang2021fair}, including concerns around group fairness~\cite{bose2019compositional,rahman2019fairwalk}, individual fairness~\cite{kang2020inform,dong2021individual}, and other task-specific fairness criteria~\cite{gupta2021correcting,jalali2023fairness,li2021dyadic}. Common approaches to achieve fairness in graphs include adding regularization~\cite{yao2017beyond,fu2020fairness}, using fairness constraints~\cite{rahman2019fairwalk}, edge rewiring~\cite{he2018hidden,jalali2023fairness}, and adversarial learning~\cite{dai2021say,wu2021learning}. These techniques are employed in a wide variety of contexts, from node classification and link prediction to community detection, among others. Fundamentally, the goal of fair graph mining is to guarantee that the insights and patterns extracted from these graph mining techniques are not only accurate but also unbiased, ensuring equitable representation and consideration of all groups within the graph's structure.

In this work, unique to previous research in fair clustering and fair graph mining, we consider a novel fairness notion in graph partition, and propose a corresponding solution to achieve balance in both node and edge. Previous work on fair spectral clustering~\cite{kleindessner2019guarantees} also pursues fairness but only in terms of the proportion of nodes in each cluster, while we consider all-around fairness in graph partition and also conduct 


\section{Notation and Problem Formulation}
\label{sec:notation}

\subsection{Preliminaries}

Let $G=(V,E,A,X)$ be an undirected graph, where $V=\{ v_{1},...,v_{n}\}$ is the node set with $E=\{e_{ij}\}_{i,j=1}^{n}$ representing edges between nodes, $A$ and $X$ is the graph adjacency matrix and nodes' attributes. Here we consider a binary adjacency matrix, i.e., if there is an edge between $v_{i}$ and $v_{j}$, then $e_{ij}=1$, otherwise $e_{ij}=0$. Each node holds a demographic attribute $S(v)$. We use $S=\{s_{1},...,s_{h}\}$ denote demographic groups such that $G=\dot{\cup}_{i \in[h]} s_{i}$, where $h$ is the number of demographic groups. Assume there exists a partition $C$ such that $G=\dot{\cup}_{j \in[k]} c_{j}$, where $C=\{c_{1},...c_{k}\}$ and $k$ is the number of clusters. $N_{j}^{i}=\{v\in c_{j} \cap s_{i}\mid |v|\}$ defines the number of nodes in cluster $c_{j}$ that also in demographic groups $s_{i}$. Based on these notations, we introduce several definitions related to node balance and edge balance in the context of fair graph partition.
\vspace{-1mm}
\begin{definition}[Node Balance \cite{chierichetti2017fair}]
\label{NB}
Given a graph $G=(V,E)$, the demographic groups $S$ and a partition $C$, the node balance (NB) can be defined as follows: 
\begin{equation} 
\mathrm{NB}\left(c_k\right)=\min _{h \neq h^{\prime} \in[h]} \frac{\left|s_h \cap c_k\right|}{\left|s_{h^{\prime}} \cap c_k\right|} \in[0,1].
\end{equation}
\end{definition}\vspace{-1mm}
Node balance essentially seeks a clustering where each group is represented in every cluster with a proportion similar to that in the entire dataset $V$. The greater the balance within each cluster, the fairer the clustering. The node balance of the clustering algorithm is the minimum node balance among each cluster. Inspired and follow up by this definition, we form our insight into edge balance. We first aim to define two distinct types of edges, with the criterion for distinguishing them being whether the two nodes connected by the edge belong to the same group.
\begin{definition}[Inter-Edge]
\label{interedge}
We define an edge between different demographic groups as an inter-edge. $IE(c_{k})$ represents the number of inter edges in cluster $c_{k}$.
\end{definition}\vspace{-1mm}

Inter-edge is the edge where nodes are from different sensitive groups. On the other side, edges formed by nodes from the same demographic groups are called {intra-edge}. One may note that we consider edges within one cluster but not edges across different clusters, as we care about inner cluster group fairness. Based on this definition, we propose a metric that measures the number of {inter-edges} to evaluate whether cluster edges are balanced, referred to as edge balance.
\begin{definition}[Edge Balance]
\label{EB}
Given a graph $G=(V,E)$, the demographic groups $S$ and a partition $C$, the edge balance (EB) of cluster $c_{k}$ is defined as follows: \vspace{-1mm}

\begin{equation}
E B\left(c_k\right)=\frac{\log \left(I E\left(c_k\right)+C\right)}{\log \left(\sum_{\substack{i < j }}^h n_k^i n_k^j\right)} \in[0,1],
\end{equation}

\end{definition}\vspace{-1mm}
where $n_k^h=1$ when $N_k^h=0$, otherwise $n_k^h=N_k^h$; $C=1$ when $I E\left(c_k\right)=0$, otherwise $C=0$. Edge balance refers to the ratio of the number of {inter-edges} present when all nodes from different demographic groups are fully connected. A cluster with nodes from different groups fully connected will have an edge balance of 1, indicating perfect balance; while a cluster without any inter-edges will have an edge balance of 0, indicating no connection between nodes of different groups.

\subsection{Problem Definition}
\label{sec:prob}
In this section, we aim to formalize our goals and illustrate the research challenges that exist under such objectives. To succinctly summarize, we characterize our problem as follows:
\begin{center}\vspace{-1mm}
\textbf{\emph{Can we generate dual node and edge fair representations while achieving balance nodes and edges distributions when performing partition with such representations?}}
\end{center}\vspace{-1mm}
In general, many existing methods achieve node fairness by pre-dividing, recombining \cite{schmidt2020fair}, or adding constraints \cite{kleindessner2019guarantees, li2020deep} to the original clustering methods. When considering edge balance, they are constrained by the original method's handling of edges. For instance, when we introduce fairness constraints into spectral clustering, we still need to use the RatioCut \cite{SC} approach for clustering, which is designed to favor cutting edges with smaller weights. Therefore, we need to design a preprocessing scheme to ensure that the embeddings we obtain are unbiased during the final partitioning. Furthermore, there are various ways in which edges are represented during the embedding process, and we hope to use a single framework to represent both nodes and edges latently.

\subsection{Physical Model Exploration}

In this section, we intend to empirically validate our motivation that partitions without edge balance concerns could occur as node balance and edge balance don't share a linear correlation. Furthermore, we demonstrate the feasibility of achieving both node and edge fairness within a single partition.

\begin{figure}[t]
    \centering
    \includegraphics[width=1.0\textwidth]{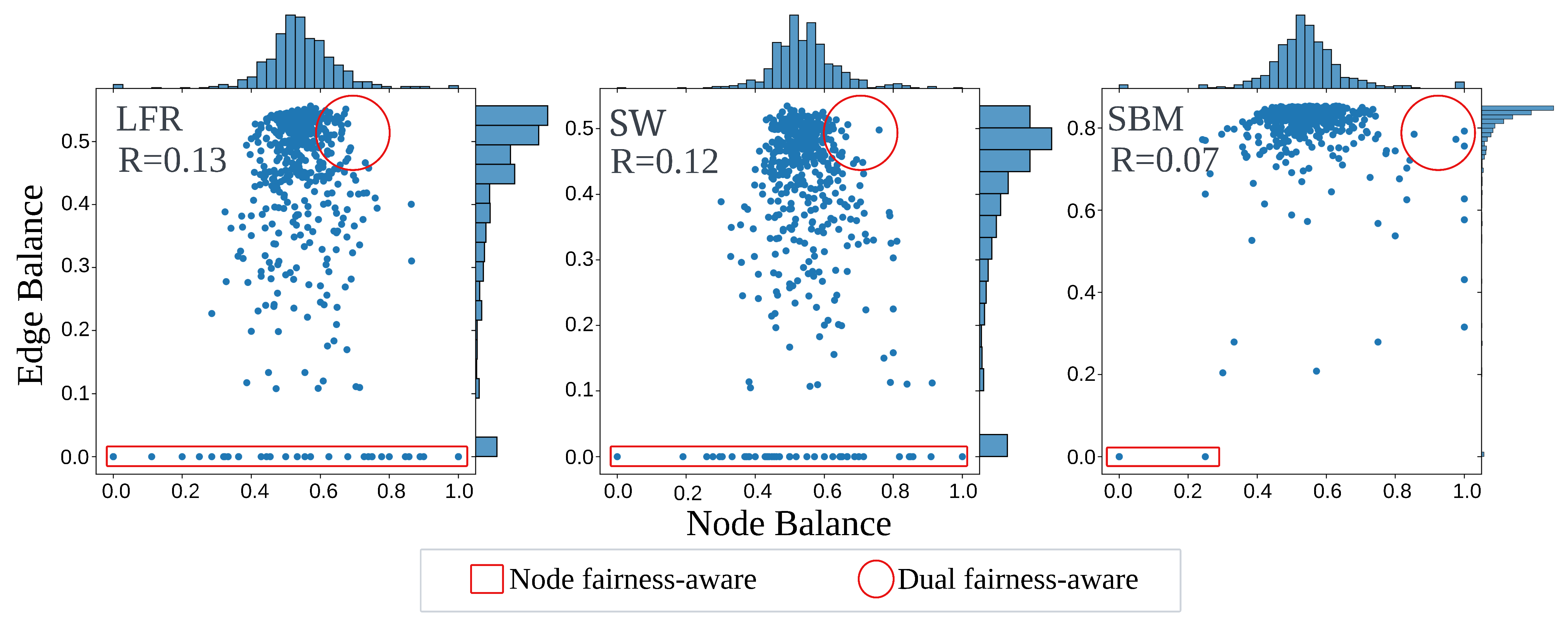}
    \caption{Random Clustering Experiment: the empirical correlation between node balance and edge balance in random clustering on three synthetic datasets. R indicates the Pearson correlation coefficient of node and edge balance.}
    \label{fig:random}
\vspace{-5mm}
\end{figure}

We conduct a random clustering experiment in Figure~\ref{fig:random} by using three popular graph generation models, namely the Lancichinetti-Fortunato-Radicchi benchmark (LFR) \cite{lancichinetti2008benchmark}, Watts-Strogatz model (SW) \cite{watts1998collective}, and Stochastic Block Model (SBM) \cite{holland1983stochastic}. We constructed graphs with 10k nodes, including two demographic groups of 5k and 5k nodes, respectively. We repeatedly perform random clustering among different clusters' sizes and report node and edge balance 500 times. 

Figure~\ref{fig:random} highlight two types of partitions: node fairness-aware and dual fairness-aware. Node fairness-aware partitions, represented by points in the plot's red square, maintain zero edge balance irrespective of node balance. These often occur in unbalanced clusters where at least one lacks edge connections between demographic groups, aligning with our concerns of partitions without edge balance constraints, as seen in Figure~\ref{fig:intro}. Additionally, we have not identified a linear relationship between node and edge balance, suggesting that constraining node balance alone is not sufficient to prevent edge disproportionality. Conversely, the ideal dual fairness-aware partition, represented by points in the red circle, can achieve high node and edge balance within a single partition.


\section{Methodology}
In this section, we provide an in-depth explanation of our suggested dual node and edge fairness framework and its technical details. We begin with an overview of the framework structure and thoroughly discuss the components and their respective objective functions. \vspace{-1mm}

\subsection{Framework Overview}
\begin{figure}[t]
\centering
\includegraphics[width=\textwidth]{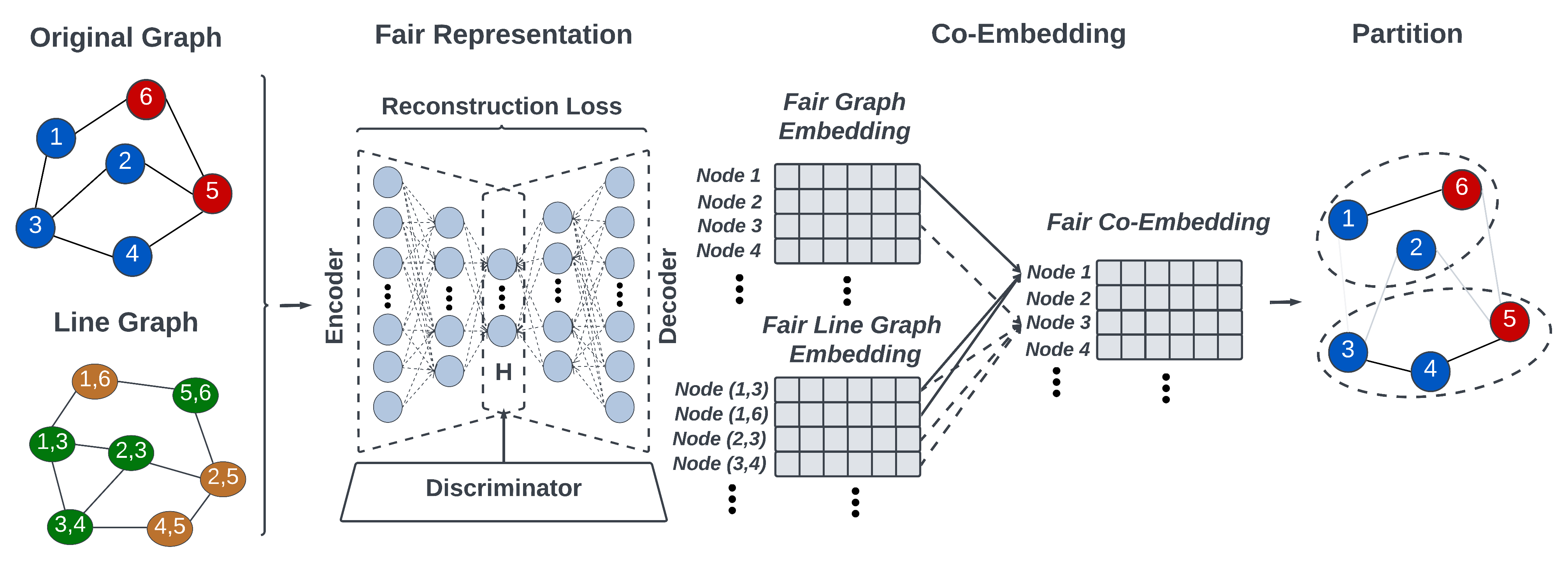}
\caption{Overview of dual fairness-aware graph partition framework. The framework incorporates adversarial learning in generating a fair co-embedding from the original graph and the line graph.}
\label{fig:framework}
\centering
\vspace{-4mm}
\end{figure}

We propose a dual fairness-aware graph partition framework to achieve fair feature representations while preserving the balance of nodes and edges demographic information. The overall framework, depicted in Figure~\ref{fig:framework}, incorporates adversarial learning for both node and edge representations to ensure fairness. To align edge representation with node representation effectively, we apply a line graph transformation, where the edge objects in the original graph are switched to node objects in the line graph. Using Graph-MLP \cite{graphMLP}, we construct a generator combined with a two-layer perceptron discriminator, facilitating the learning process while preventing messages from passing between nodes and edges. Both input graphs share the same construction for hidden demographic information. Finally, the two post-processed fair representations are combined using the node and node pair index in the line graph assignment to generate a cooperative fair representation for partition.


\subsection{Line Graph Transformation}
A line graph \cite{harary1960some} is a new graph that represents the edges of a given graph as nodes, with edges connecting adjacent edges in the original graph. The line graph is commonly used in graph mining learning, especially in edge representation learning \cite{jiang2019censnet,chen2019gl2vec,evans2010line,choudhary2021atomistic}.
This transformation captures higher-order connectivity patterns between edges in the original graph and allows the use of well-established techniques and algorithms designed for node representation learning. 

Formally, suppose we have a graph $G=(V^G,E^G,A^G,X^G)$, the adjacency information can be represent as $A^G_{ij}=1$ when $ij\in E^G,$ and 0 otherwise. A line graph $L=(V^L,E^L,A^L,X^L)$ can be constructed by considering the edges in $G$ as nodes in $L$: $V^L=E^G$, and the edges set and adjacency information could be mapping as:
\begin{equation}
E^L=\{[(i j),(j k)]|(i j),(j k) \in E^G\}, \text{and }A_{i j,kl}^L=\left\{\begin{array}{ll}
1 & [(i j),(kl)] \in E^L \\
0 &  \text{  Others }
\end{array}, \right.
\end{equation}
where $i,j,k,l\in V^G$. Based on this adjacency mapping, we can write the statistics as follows:
\begin{equation}
\left|V^L\right|=\left|E^G\right|, \text{and} \left|E^L\right|=\sum_{i=1}^n \frac{d_i\left(d_{i}-1\right)}{2},
\end{equation}
where $d_{i}$ is the degree number of each node in $V$ and $n=\left|V^G\right|$.  
From the view of demographic information of this transformation, the original graph contains nodes from different demographic groups, and edges connecting these groups are {inter-edges}. When we map it into the line graph, these edge divisions will become the new demographic information for the nodes in the line graph. In this case, we convert the demographic information of edges in the original graph to the nodes' group difference in the line graph. 
Let $\mathcal{F}$ and $\mathcal{D}$ represent the functions of the encoder and discriminator. $G$ and $L$ share the same adversarial learning structure but trained separately; two fair hidden embeddings are generated by: ${H^G}=\mathcal{D} \circ \mathcal{F}(X^G)$ and ${H^L}=\mathcal{D} \circ\mathcal{F}(X^L)$. Each item in $H^G$ stands for a node $i$ representation of $V^G$, and each element in $H^L$ serves a node $ij$ for $V^L$, where $i$ and $ij$ follows a node to node pair index mapping. The final co-embedding can be expressed as: 
\begin{equation}
H_i=H_i^G+\frac{1}{n} \sum_{j \neq i}^n H_{i j}^L\text{, where } i\in V^G \text{and } ij\in V^L.
\end{equation}\vspace{-6mm}

\subsection{Objective Function}
The objective function of the proposed dual fairness-aware framework consists of the latent representation reconstruction loss and the fairness-adversarial learning loss. In the subsequent paragraphs, we offer detailed information on these distinct components employed in our proposed model from the view of the input original graph $G$, while noting that $L$ remains the same.
\vspace{-3mm}

\paragraph{Reconstruction Loss.}
Graph reconstruction loss trains our encoder $\mathcal{F}$ to learn a node representation that preserves the graph's structure. The generator $\mathcal{F}$ takes $G=(V,E,A,X)$ as input and performs an unsupervised reconstruction learning on attribute matrix $A$. Our framework uses Graph-MLP \cite{graphMLP} for constructing an encoder while avoiding message passing during the aggregation compared to other structures. Graph-MLP uses three perceptron layers for learning node representation optimized with Neighbouring Contrastive(Ncontrast) Loss. Ncontrast loss ensures that connected nodes are similar while unconnected nodes are distant in the feature space. We use a 2-hop adjacency matrix to consider nodes' similarity while the loss promotes nodes within a 2-hop distance closer and pushing others further away. We also use Mean Square Error (MSE) loss to facilitate feature extraction incorporating graph connection information. The objective function of reconstruction $\mathcal{L}_R$ is:
\begin{equation}
\mathcal{L_R}=\frac{1}{|V|} \sum_{v_i \in V}\left(X_i-\mathcal{F}\left(X_i\right)\right)^2-\log \frac{\sum_{v_{i}, v_{j} \in V} {A}^2_{i j}\exp\left[\operatorname{sim} \left(\mathcal{F}\left(X_i\right), \mathcal{F}\left(X_j\right)\right)\right]}{\sum_{v_{i}, v_{k} \in V} \exp\left[\operatorname{sim} \left(\mathcal{F}\left(X_i\right), \mathcal{F}\left(X_k\right)\right) \right]},
\end{equation}
where sim denotes the cosine similarity and ${A}^2$ represent a 2-hop adjacency matrix of graph $G$.
\vspace{-3mm}
\paragraph{Fairness-Adversarial Loss.}
Adversarial learning has been demonstrated to reduce bias and discrimination in representations effectively. For categorical demographic groups, our goal is to attain fairness by promoting partition balance group distribution in each cluster. Generally, an adversary predicts demographic information based on learned representations, and simultaneously, the encoder is trained to learn representations that make it challenging for the adversary to predict correctly. In our task, the fairness-adversarial loss can be written as: 
\begin{equation}
\mathcal{L}_A=\frac{1}{|V|} \sum_{v_i \in V}\left[\mathbb{E}_{X_i \sim p(X_i \mid s=1)} \log (\mathcal{D} \circ \mathcal{F}(X_i))+\mathbb{E}_{X_i \sim p(X_i \mid s=0)} \log (\mathcal{D} \circ \mathcal{F}(X_i))\right],
\end{equation}
where $s$ is the demographic group of node $v_i$ and $\circ$ is the function composition. 

For the last clustering algorithm, we use the Spectral Clustering \cite{SC} algorithm to perform on our cooperative fair representations. To sum up, the overall objective function for our dual node and edge fairness-aware partition framework can be written as the following minimax game:
\begin{equation}
\begin{aligned}
& \min _{\theta_R} \max _{\theta_A} \mathcal{L}_R-\mathcal{L}_A, \\
\end{aligned}
\end{equation}
$\theta_R$, $\theta_A$ denotes parameters in the learning process. To this end, we employ a min-max game strategy to generate fair representations, with the encoder producing the desired representation for capturing graph information and the discriminator achieving the fairness process. In the case of the original graph, fair representations are attained when the discriminator is unable to differentiate between representations originating from various protected subgroups; similarly, for the transform line graph, fair representations occur when the discriminator cannot distinguish representations of nodes in the line graph are formed by {inter-edges} or by {intra-edges} of the original graph.


\section{Experiment}
\subsection{Experimental Setup}
\paragraph{Datasets.}\vspace{-1mm}
We choose five widely used social networks datasets in the fair graph mining area to conduct our experiments, which include \textit{NBA} \cite{dai2021say}, \textit{Friendship}, \textit{Facebook} \cite{highschool}, Oka97, and UNC28 \cite{red2011comparing}. The \textit{NBA} dataset incorporates 400 NBA basketball players along with their performance statistics from the 2016-2017 seasons. Links between each player are established based on their social media interactions. The dataset's demographic information is the nationality of each player, segregated into U.S. players and overseas players. The \textit{Friendship} and \textit{Facebook} datasets depict contact and Facebook following connections among students at a high school in Marseilles, France, as of December 2013. We employ the students' gender as the demographic information. \textit{Ok97} and \textit{UNC28} represent two high school social networks. Connections in these networks indicate friendships between users, and the demographic groups are classified based on gender. The dataset statistics are summarized in Table~\ref{tab:sta}. Here, \#Group refers to the quantity of demographic groups, \#Class signifies the count of ground-truth labels, \#Inter and \#Intra denotes the total number of inter-edges and intra-edges as explained in Section~\ref{sec:notation}.  Group Ratio is the ratio of two demographic groups in each dataset. Inter and intra ratio denotes the proportion between the actual number of inter and intra edges to the potential number of links that would exist if the graph were fully connected. These two variables illustrate the density of inter and intra edges connections in the graph.

\paragraph{Competitive Methods.}\vspace{-1mm}
Based on our unsupervised scenario, we include two fair graph mining methods, Fair Spectral Clustering (FairSC) \cite{kleindessner2019guarantees} and FairWalk \cite{rahman2019fairwalk}. Fair Spectral Clustering considers fairness as a linear constraint on graph laplacian matrix, and this constraint guides SC to find a fair clustering if such a one exists. Fairwalk introduces a fairness-aware transition probability distribution to balance the random walks. This modified probability distribution ensures that the generated walks give equal representation to different groups of nodes while maintaining the structural properties of the original graph. We also include four partitions and graph representation learning methods including K-means \cite{Kmeans}, Spectral Clustering (SC) \cite{SC}, GCN \cite{kipf2016semi}, and GAT \cite{velickovic2017graph}. 

\paragraph{Implementation.}\vspace{-1mm}
We implement our framework using PyTorch \cite{paszke2019pytorch} on Python3.6 with the NVIDIA GTX 3080 GPU and 64GB RAM. We use node2vec \cite{grover2016node2vec} to generate line graphs' node attributes. The encoder consists of three layers with batch normalization and dropout, and the decoder reverses the architecture. We run the experiment 20 times and report the mean value with the standard deviation. 

\begin{table}[t]
\centering
\caption{Statistic for datasets in experiments}
\label{tab:sta}
\newcolumntype{Y}{>{\centering\arraybackslash}X}
\begin{tabularx}{\columnwidth}{@{}YYYYYYccc@{}}
\toprule
Datasets            & \# Node & \# Edge  & \# Class & \# Inter & \# Intra & Group Ratio  & Inter Ratio   & Intra Ratio                    \\ \midrule
\textit{NBA}        & 400    & 10,621 & 3              & 2935         & 7,686       & 0.361       & 9.32e-2       & 15.8e-2                                    \\
\textit{Friendship} & 134    & 406    & 4              & 44           & 362         & 0.675       & 1.01e-2       & 7.88e-2                                  \\
\textit{Facebook}   & 156    & 1437   & 4              & 170          & 1,267       & 0.835       & 2.81e-2       & 20.9e-2                                 \\
\textit{Ok97}       & 3,111   & 73,320 & 5             & 26,862       & 46,458      & 0.977       & 1.11e-2         &1.92e-2                                  \\
\textit{UNC28}      & 4,018   & 65,287 & 5             & 29,075       & 36,212      & 0.730       & 7.38e-3         &8.76e-3                                     \\ \bottomrule
\end{tabularx}
\vspace{-4mm}
\end{table}

\begin{table}[t]
\centering
\small
\caption{Quantitative results of fair graph partition experiment on five datasets}
\label{table:partition}
\newcolumntype{Y}{>{\centering\arraybackslash}X}
\begin{tabularx}{1\columnwidth}{@{}ccYYYYYYY@{}}
\toprule
{Datasets}   & {Metric}                             & {Ours}        & {FairSC}     & {Fairwalk}    & {K-Means}     & {SC} & {GCN}         & {GAT}         \\ \midrule
\multirow{ 5}{*}{\textit{NBA}}          & {ACC}           & {.404 {\tiny± .025}} & {.413 {\tiny± .001}}   & \textbf{.468 {\tiny± .016}} & {.443 {\tiny± .035}} & {.373 {\tiny± .014}} & {.396 {\tiny± .003}} & {.392 {\tiny± .017}} \\
         & {Min/NB} & \textbf{.332 {\tiny± .016}} & {.050 {\tiny± .021}} & {.302 {\tiny± .032}} & {.136 {\tiny± .112}} & {.282 {\tiny± .042}}  & {.319 {\tiny± .015}} & {.225 {\tiny± .102}} \\
         & {Mean/NB} & \textbf{.384 {\tiny± .010}} & {.119 {\tiny± .021}} & {.360 {\tiny± .011}} & {.311 {\tiny± .058}} & {.351 {\tiny± .042}}  & {.347 {\tiny± .003}} & {.309 {\tiny± .064}} \\
           & {Min/EB} & \textbf{.653 {\tiny± .040}} & {.102 {\tiny± .132}} & {.119 {\tiny± .102}} & {.353 {\tiny± .180}} & {.588 {\tiny± .028}} & {.602 {\tiny± .002}} & {.410 {\tiny± .201}}  \\  
           & {Mean/EB} & \textbf{.709 {\tiny± .006}} & {.258 {\tiny± .032}} & {.515 {\tiny± .032}} & {.616 {\tiny± .083}} & {.655 {\tiny± .005}} & {.688 {\tiny± .004}} & {.542 {\tiny± .163}}  \\ \midrule

\multirow{ 5}{*}{\textit{Friendship}}   & {ACC}  & {.523 {\tiny± .083}} & {.396 {\tiny± .003}}   & {.342 {\tiny± .027}} & \textbf{.690 {\tiny± .057}} & {.582 {\tiny± .006}} & {.403 {\tiny± .031}} & {.418 {\tiny± .049}} \\
                                        & {Min/NB} & \textbf{.472 {\tiny± .061}} & {.132 {\tiny± .034}} & {.371 {\tiny± .102}} & {.116 {\tiny± .078}} & {.085 {\tiny± .015}}  & {.275 {\tiny± .130}} & {.231 {\tiny± .097}} \\
                                        & {Mean/NB} & \textbf{.687 {\tiny± .034}} & {.175 {\tiny± .021}} & {.630 {\tiny± .041}} & {.523 {\tiny± .058}} & {.310 {\tiny± .091}}  & {.601 {\tiny± .021}} & {.501 {\tiny± .007}} \\
                                        & {Min/EB} & \textbf{.359 {\tiny± .091}} & {.132 {\tiny± .002}} & {.106 {\tiny± .112}} & {.231 {\tiny± .069}} & {.072 {\tiny± .004}} & {.208 {\tiny± .165}} & {.074 {\tiny± .130}}  \\ 
                                         & {Mean/EB} & {.443 {\tiny± .031}} & {.155 {\tiny± .012}} & {.339 {\tiny± .037}} & \textbf{.485 {\tiny± .018}} & {.318 {\tiny± .002}} & {.413 {\tiny± .004}} & {.376 {\tiny± .074}} \\ \midrule

\multirow{5}{*}{\textit{Facebook}}     & {ACC}           & {.644 {\tiny± .064}} & {.701 {\tiny± .013}}   & \textbf{.357 {\tiny± .024}} & {.782 {\tiny± .005}} & {.767 {\tiny± .009}} & {.351 {\tiny± .029}} & {.399 {\tiny± .041}} \\
                                        & {Min/NB} & \textbf{.569 {\tiny± .072}} & {.080 {\tiny± .002}} & {.502 {\tiny± .021}} & {.333 {\tiny± .02}} & {.276 {\tiny± .022}}  & {.421 {\tiny± .053}} & {.476 {\tiny± .085}} \\
                                        & {Mean/NB} & {.745 {\tiny± .063}} & {.270 {\tiny± .021}} & {.730 {\tiny± .007}} & \textbf{.812 {\tiny± .021}} & {.610 {\tiny± .091}}  & {.631 {\tiny± .018}} & {.643 {\tiny± .007}} \\
                                        & {Min/EB} & \textbf{.574 {\tiny± .086}} & {.060 {\tiny± .003}} & {.039 {\tiny± .099}} & {.470 {\tiny± .004}} & {.518 {\tiny± .011}} & {.542 {\tiny± .035}} & {.464 {\tiny± .097}}  \\   
                                        & {Mean/EB} & \textbf{.610 {\tiny± .042}} & {.396 {\tiny± .012}} & {.49 {\tiny± .022}} & {.582 {\tiny± .018}} & {.537 {\tiny± .034}} & {.579 {\tiny± .015}} & {.581 {\tiny± .024}} \\\midrule

\multirow{5}{*}{\textit{Ok97}}     & {ACC}         & {.275 {\tiny± .064}} & {.302 {\tiny± .021}}   & {.253 {\tiny± .004}} & {.262 {\tiny± .005}} & {.284 {\tiny± .008}} & {.280 {\tiny± .037}} & \textbf{.374 {\tiny± .028}} \\
                                    & {Min/NB} & {.585 {\tiny± .021}} & {.136 {\tiny± .062}} & {.521 {\tiny± .013}} & \textbf{.708 {\tiny± .007}} & {.428 {\tiny± .001}}  & {.471 {\tiny± .038}} & {.452 {\tiny± .218}} \\
                                    & {Mean/NB} & {.829 {\tiny± .012}} & {.439{\tiny± .025}} & {.599 {\tiny± .014}} & \textbf{.841 {\tiny± .025}} & {.788 {\tiny± .008}}  & {.656 {\tiny± .029}} & {.771 {\tiny± .069}} \\
                                    & {Min/EB} & \textbf{.401 {\tiny± .004}} & {.116 {\tiny± .072}} & {.349 {\tiny± .009}} & {.346 {\tiny± .040}} & {.043 {\tiny± .009}} & {.068 {\tiny± .086}} & {.258 {\tiny± .045}}  \\
                                    & {Mean/EB} & \textbf{.593 {\tiny± .012}} & {.369{\tiny± .049}} & {.581 {\tiny± .002}} & {.581 {\tiny± .004}} & {.285 {\tiny± .006}} & {.522 {\tiny± .028}} & {.495 {\tiny± .036}} \\\midrule

\multirow{ 5}{*}{\textit{UNC28}}    & {ACC} & {.299 {\tiny± .003}} & {.333 {\tiny± .004}}  & \textbf{.324 {\tiny± .012}} & {.270 {\tiny± .004}} & {.324 {\tiny± .011}} & {.292 {\tiny± .020}} & {.411 {\tiny± .037}} \\
                                    & {Min/NB} & \textbf{.654 {\tiny± .053}} & {.253 {\tiny± .041}} & {.525 {\tiny± .052}} & {.587 {\tiny± .006}} & {.632 {\tiny± .014}}  & {.598 {\tiny± .012}} & {.641 {\tiny± .022}} \\
                                    & {Mean/NB} & \textbf{.723 {\tiny± .012}} & {.629{\tiny± .031}} & {.701 {\tiny± .014}} & {.712 {\tiny± .005}} & {.697 {\tiny± .027}}  & {.686 {\tiny± .010}} & {.655 {\tiny± .016}} \\
                                    & {Min/EB} & \textbf{.525 {\tiny± .040}} & {.064 {\tiny± .028}} & {.101 {\tiny± .095}} & {.501 {\tiny± .007}} & {.138 {\tiny± .060}} & {.242 {\tiny± .002}} & {.221 {\tiny± .014}}  \\  
                                    & {Mean/EB} & \textbf{.590 {\tiny± .020}} & {.130{\tiny± .012}} & {.343 {\tiny± .002}} & {.560 {\tiny± .022}} & {.458 {\tiny± .018}} & {.546 {\tiny± .015}} & {.509 {\tiny± .013}} \\
\bottomrule
\end{tabularx}
\vspace{-4mm}
\end{table}
\paragraph{Metrics.} \vspace{-1mm}
We conduct five measurements in the fair graph partition experiment and its in-depth exploration section. For the fair graph partition experiment, we use clustering accuracy (ACC) by comparing the attained partition and external ground truth labels to measure the clustering utility; Node Balance (NB, Definition~\ref{NB}) and Edge Balance (EB, Definition~\ref{EB}) to measure the node distribution and edge connection balance among the clusters. To better capture the fairness information within a partition, we consider reporting the minimum and the mean balance value in each partition cluster. In general, the min balance value follows our definition in Section~\ref{sec:notation}, and the mean value relaxes the constraint to average. The higher NB and EB, the fairer the algorithm is. ACC is calculated as follows:
\begin{equation} 
\text{ACC}=\frac{1}{n} \sum_{i=1}^n\left(\operatorname{Perm}\left(\hat{y}_i\right)=y_i\right),
\end{equation}
where $y \in\{0,1\}$ denotes the binary ground truth label and $\hat{y} \in\{0,1\}$ denotes the prediction of the classifier. Perm is a permutation mapping function that maps each cluster $\hat{y}_i$ label to the ground truth label $y_i$ and $n$ is the total data instance number.

In the fair node classification task, we introduce two popular used node fairness metrics: Statistical Parity($\Delta_{S P}$) \cite{beutel2017data} and Equal Opportunity($\Delta_{E O}$) \cite{louizos2015variational}. $\Delta_{S P}$ necessitates that predictions remain uncorrelated to the demographic information, while $\Delta_{E O}$ demands that the classifier maintains consistent true positive rates among all subgroups. The lower the value of these two measures, the more equitable the classifier becomes. These metrics are calculated as follows:
\begin{equation}
\Delta_{S P}=|P(\hat{y}=1 \mid s=0)-P(\hat{y}=1 \mid s=1)| \text {, }
\end{equation}
\begin{equation}
\Delta_{E O}=|P(\hat{y}=1 \mid y=1, s=0)-P(\hat{y}=1 \mid y=1, s=1)|,
\end{equation}
where $y \in\{0,1\}$ denotes the binary label and $\hat{y} \in\{0,1\}$ denotes the prediction of the classifier. $s \in\{0,1\}$ presents the demographic group. 
In the fair link prediction experiment, we use Network Modularity($Q$) \cite{newman2004finding} and \textit{Modred} \cite{masrour2020bursting} to measure the fairness of adding predicted edges into the graph. Intuitively, we target the fairness of  the formation of links. Modred \cite{masrour2020bursting} assesses the decrease in modularity metric to ascertain if the modified network, derived from link prediction outcomes, exhibits a bias toward generating a higher number of inter-group or intra-group connections. A positive Modred value signifies that the link prediction algorithm forecasts more inter-group connections than the ground truth network. Conversely, a negative value implies that the algorithm predicts a higher volume of intra-group links than the actual network. In this case, we use the sum of $|\text{Modred}|$ of all clusters among the partition to measure the fairness of an algorithm. These metrics are calculated as follows:
\begin{equation}
Q=\frac{1}{2 m} \sum_{i j}\left(A_{i j}-\frac{d_i d_j}{2 m}\right) \delta\left(c_i, c_j\right), \text{and } |\text{Modred}|=|\frac{Q_{\mathrm{gt}}-Q_{\mathrm{pred}}}{Q_{\mathrm{gt}}}|.
\end{equation}
$A$ is the adjacency matrix of the network, $\delta\left(c_i, c_j\right)$ is the Kronecker delta function, $c_i$ is the community of node $i$, and $d_i$ is its corresponding degree and $m$ is total number of edges. $Q_{\mathrm{gt}}$ denotes the modularity measure of the network with ground truth label, and $Q_{\mathrm{pred}}$ represents the modularity metric of the network formed by adding the predicted links to the initial network.

\subsection{Algorithmic Performance}
We conduct experiments of our dual node and edge fairness-aware partition framework on all five datasets. As we illustrate in Section~\ref{sec:prob}, our motivation is to achieve balance node and edge distribution simultaneously in a single partition. We transform a line graph for each input data and feed them into the same adversarial learning structure. After constructing the fair co-embedding representations, we use spectral clustering as our final cluster assignment method. 

Table~\ref{table:partition} presents a quantitative evaluation of our framework on graph partition, where the highest value in each row is indicated in bold. The fairness evaluation reveals that conventional graph partition methods exhibit poor fairness quality in disregarding edge fairness. In comparison, our framework outperforms existing fair partition methods in both node balance and edge balance. On average, our method achieves an 18.6\% improvement in node balance compared to other fair methods, and it also demonstrates the highest increase in edge balance utility across most datasets. We observe that some methods may produce a low minimum balance value but a relatively high mean balance value, indicating that at least one cluster in the partition has a fully unbalanced distribution, and our framework avoids this situation. While we do notice a performance gap in node balance for the \textit{Ok97} dataset, our method achieves the highest sum of node and edge balance among all other methods. This measurement aligns with our motivation to consider dual node and edge balance as a unified objective in fair graph mining. Regarding partition accuracy, we observe a trade-off between fairness and clustering benefits, as fair methods result in a utility loss compared to unconstrained methods.


\subsection{In-depth Exploration}
Despite our high-quality performance on fair graph partitions, our objective extends beyond facilitating an unsupervised partitioning process. We aim to ensure that our model also contributes to enhancing fairness in downstream tasks, effectively reducing and even eradicating data bias in applications. Essentially, our framework can function as a preprocessing method, utilizing the partition labels derived from our model to guide message aggregation. 


\paragraph{Node Classification.}\vspace{-2mm}
We first conduct a node classification task, where our framework serves as a pseudo-label generator. In this experiment, we first obtain the pseudo labels of corresponding clusters using our framework, then feed the pseudo labels to train a GCN classifier, and finally evaluate this classifier on the output data of other testing graphs. In Table~\ref{tab:classification}, we present the fairness performance of the GCN trained on the pseudo labels generated by the oka and unc datasets, respectively, when performing node classification on the remaining datasets. The classifier trained with the pseudo labels generated by our framework exhibits lower unfairness when performing node classification on entirely new graphs. We also observed that the classifier is sensitive to the input training graphs. Classifiers trained on different input graphs exhibit varying fairness performance for the same test set.

\begin{table}[t]
\centering
\small
\caption{Fair node classification results of GNN trained on partition labels by our framework}
\label{tab:classification} 
\newcolumntype{Y}{>{\centering\arraybackslash}X}
\begin{tabularx}{\columnwidth}{@{}YYcccccccc@{}}
\toprule
\multirow{2}{*}{Training} & \multirow{2}{*}{Testing} & \multicolumn{4}{c}{$\Delta_{SP}(\%)$}   & \multicolumn{4}{c}{$\Delta_{EO}(\%)$}   \\ \cmidrule(lr){3-6}\cmidrule(l){7-10}
                       &                       & Ours & FairSC & Fairwalk & K-means & Ours & FairSC & Fairwalk & K-means \\ \midrule
\multirow{3}{*}{\textit{Ok97}}  & \textit{NBA}     &\textbf{0.8 {\tiny± 0.2}}     & {3.8 {\tiny± 1.3}}    & {1.8 {\tiny± 1.1}}   &{4.4 {\tiny± 1.7}}      &\textbf{3.4 {\tiny± 1.4}}  & {4.6 {\tiny± 3.0}}   & {9.8 {\tiny± 1.9}}   &   {6.1 {\tiny± 2.3}}   \\
                       & \textit{Friendship}&\textbf{6.5 {\tiny± 0.8}}      &{8.2 {\tiny± 1.6}}        &{8.0 {\tiny± 1.4}}   &{10.0 {\tiny± 2.7}}       &\textbf{9.5 {\tiny± 4.4}}      &{13.1 {\tiny± 7.4}}        &{20.1 {\tiny± 7.3}} &{9.8 {\tiny± 1.3}}\\
                       & \textit{Facebook}  &\textbf{0.9 {\tiny± 1.2}}      &{5.2 {\tiny± 2.4}}        &{4.9 {\tiny± 3.3}}   &{3.3 {\tiny± 1.2}}        &\textbf{11 {\tiny± 7.3}}       &{24.1 {\tiny± 9.8}}        &{18.7 {\tiny± 8.1}}  &{13.2 {\tiny± 2.1}}           \\ \midrule
\multirow{3}{*}{\textit{UNC28}} & \textit{NBA}&\textbf{0.1 {\tiny± 0.07}} & {2.5 {\tiny± 1.7}}  & {3.3 {\tiny± 2.2}} & {1.0 {\tiny± 0.3}} & \textbf{7.1 {\tiny± 2.7}}& {12.2 {\tiny± 4.2}}&{12.2 {\tiny± 4.2}}  &{16.2 {\tiny± 9.5}}   \\
                       & \textit{Friendship}&{6.7 {\tiny± 3.3}} & \textbf{1.6 {\tiny± 1.6}}  & {6.8 {\tiny± 2.3}} & {2.3 {\tiny± 1.3}} & {10.1 {\tiny± 2.2}}& \textbf{9.2 {\tiny± 5.2}}&{14.3 {\tiny± 7.2}}  &{11.8 {\tiny± 5.1}}\\
                       & \textit{Facebook} &\textbf{3.8 {\tiny± 1.2}} & {5.1 {\tiny± 2.3}}  & {7.8 {\tiny± 3.3}} & {15.3 {\tiny± 2.2}} & \textbf{6.7 {\tiny± 1.6}}& {8.4 {\tiny± 4.2}}&{14.1 {\tiny± 1.3}}  &{14.3 {\tiny± 8.1}} \\ \bottomrule
\end{tabularx}
\vspace{-4mm}
\end{table}

\begin{figure}[t]
    \centering
    \includegraphics[width=0.95\textwidth]{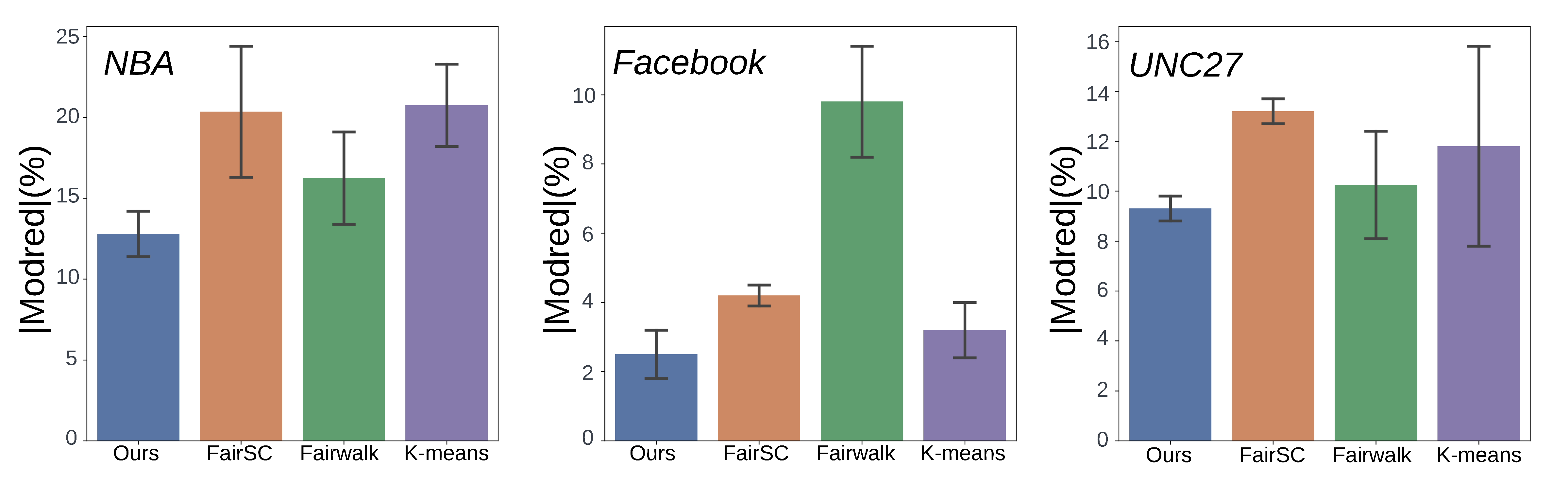}
    \caption{Experimental results of fair link prediction on three datasets}
    \label{fig:link}
    \vspace{-2mm}
\end{figure}

\paragraph{Link Prediction.}\vspace{-2mm}
In this experiment, we aim to measure whether the clusters generated by our model can be fair at the edge level in subsequent application scenarios of edge formation, such as recommendation systems. We examine whether the link prediction algorithm will bias inter-group or intra-group edges during formation. This bias can be measured by calculating the change in modularity. We first employ our model to obtain a partition and then apply the Adamic-Adar coefficient \cite{adamic2003friends} for link prediction on each cluster within that partition. We add 15\% of the high-scoring potential links to each cluster and then calculate the change in Modred for that cluster, summing their absolute values. Our framework does not add or remove edges on the original graph; the change in $Modred$ value depends on the cluster assignment of each node. In other words, we use the pseudo-label information obtained through our framework to achieve a fair link prediction similar to the node classification task. As shown in Figure~\ref{fig:link}, our method achieve the smallest sum of $|Modred|$ in these datasets. We demonstrate that the partition obtained through our method alleviates the bias towards inter-group and intra-group edges in link prediction compared to other approaches. 


\section{Conclusion}
In this paper, we explore the concept of fairness in graph partitioning from both the node and the edge perspective. We empirically validate and confirmed the feasibility of achieving dual node and edge fairness-aware partition by using three synthetic datasets. Furthermore, we introduce a co-embedding framework that facilitates learning unbiased representations, effectively concealing demographic information for graph partition. Comprehensive experimentation across several social networks revealed that our proposed method outperforms existing graph partitioning algorithms with fairness consideration. Moreover, we demonstrate that our fair partitioning can be utilized as pseudo labels to guide the following node classification and link prediction algorithm to behave fairly.

\newpage
\bibliographystyle{plain}  
\bibliography{references}  

\begin{thebibliography}{10}

\bibitem{adamic2003friends}
Lada~A Adamic and Eytan Adar.
\newblock Friends and neighbors on the web.
\newblock {\em Social networks}, 25(3):211--230, 2003.

\bibitem{ahmadian2020fair}
Sara Ahmadian, Alessandro Epasto, Marina Knittel, Ravi Kumar, Mohammad Mahdian,
  Benjamin Moseley, Philip Pham, Sergei Vassilvitskii, and Yuyan Wang.
\newblock Fair hierarchical clustering.
\newblock In {\em Advances in Neural Information Processing Systems}, 2020.

\bibitem{Kmeans}
David Arthur and Sergei Vassilvitskii.
\newblock k-means++: The advantages of careful seeding.
\newblock Technical report, Stanford, 2006.

\bibitem{backurs2019scalable}
Arturs Backurs, Piotr Indyk, Krzysztof Onak, Baruch Schieber, Ali Vakilian, and
  Tal Wagner.
\newblock Scalable fair clustering.
\newblock In {\em International Conference on Machine Learning}, 2019.

\bibitem{bedi2016community}
Punam Bedi and Chhavi Sharma.
\newblock Community detection in social networks.
\newblock {\em Wiley Interdisciplinary Reviews: Data Mining and Knowledge
  Discovery}, 6(3):115--135, 2016.

\bibitem{bera2019fair}
Suman Bera, Deeparnab Chakrabarty, Nicolas Flores, and Maryam Negahbani.
\newblock Fair algorithms for clustering.
\newblock In {\em Advances in Neural Information Processing Systems},
  volume~32, 2019.

\bibitem{beutel2017data}
Alex Beutel, Jilin Chen, Zhe Zhao, and Ed~H Chi.
\newblock Data decisions and theoretical implications when adversarially
  learning fair representations.
\newblock {\em arXiv preprint arXiv:1707.00075}, 2017.

\bibitem{bose2019compositional}
Avishek Bose and William Hamilton.
\newblock Compositional fairness constraints for graph embeddings.
\newblock In {\em International Conference on Machine Learning}, 2019.

\bibitem{chakrabarty2021better}
Deeparnab Chakrabarty and Maryam Negahbani.
\newblock Better algorithms for individually fair $ k $-clustering.
\newblock {\em arXiv preprint arXiv:2106.12150}, 2021.

\bibitem{chen2019gl2vec}
Hong Chen and Hisashi Koga.
\newblock Gl2vec: Graph embedding enriched by line graphs with edge features.
\newblock In {\em International Conference of Neural Information Processing},
  pages 3--14, 2019.

\bibitem{chhabra2022robust}
Anshuman Chhabra, Peizhao Li, Prasant Mohapatra, and Hongfu Liu.
\newblock Robust fair clustering: A novel fairness attack and defense
  framework.
\newblock {\em arXiv preprint arXiv:2210.01953}, 2022.

\bibitem{chierichetti2017fair}
Flavio Chierichetti, Ravi Kumar, Silvio Lattanzi, and Sergei Vassilvitskii.
\newblock Fair clustering through fairlets.
\newblock In {\em Advances in Neural Information Processing Systems}, 2017.

\bibitem{choudhary2021atomistic}
Kamal Choudhary and Brian DeCost.
\newblock Atomistic line graph neural network for improved materials property
  predictions.
\newblock {\em npj Computational Materials}, 7(1):185, 2021.

\bibitem{dai2021say}
Enyan Dai and Suhang Wang.
\newblock Say no to the discrimination: Learning fair graph neural networks
  with limited sensitive attribute information.
\newblock In {\em International Conference on Web Search and Data Mining},
  2021.

\bibitem{dong2021individual}
Yushun Dong, Jian Kang, Hanghang Tong, and Jundong Li.
\newblock Individual fairness for graph neural networks: A ranking based
  approach.
\newblock In {\em International Conference on Knowledge Discovery and Data
  Mining}, 2021.

\bibitem{dong2023fairness}
Yushun Dong, Jing Ma, Song Wang, Chen Chen, and Jundong Li.
\newblock Fairness in graph mining: A survey.
\newblock {\em IEEE Transactions on Knowledge and Data Engineering}, 2023.

\bibitem{esmaeili2021fair}
Seyed Esmaeili, Brian Brubach, Aravind Srinivasan, and John Dickerson.
\newblock Fair clustering under a bounded cost.
\newblock In {\em Advances in Neural Information Processing Systems}, 2021.

\bibitem{evans2010line}
Tim~S Evans and Renaud Lambiotte.
\newblock Line graphs of weighted networks for overlapping communities.
\newblock {\em The European Physical Journal B}, 77:265--272, 2010.

\bibitem{fu2020fairness}
Zuohui Fu, Yikun Xian, Ruoyuan Gao, Jieyu Zhao, Qiaoying Huang, Yingqiang Ge,
  Shuyuan Xu, Shijie Geng, Chirag Shah, Yongfeng Zhang, et~al.
\newblock Fairness-aware explainable recommendation over knowledge graphs.
\newblock In {\em International Conference on Research and Development in
  Information Retrieval}, pages 69--78, 2020.

\bibitem{grover2016node2vec}
Aditya Grover and Jure Leskovec.
\newblock node2vec: Scalable feature learning for networks.
\newblock In {\em International Conference on Knowledge Discovery and Data
  Mining}, 2016.

\bibitem{gupta2021correcting}
Shantanu Gupta, Hao Wang, Zachary Lipton, and Yuyang Wang.
\newblock Correcting exposure bias for link recommendation.
\newblock In {\em International Conference on Machine Learning}, 2021.

\bibitem{harary1960some}
Frank Harary and Robert~Z Norman.
\newblock Some properties of line digraphs.
\newblock {\em Rendiconti Del Circolo Matematico Di Palermo}, 9:161--168, 1960.

\bibitem{he2018hidden}
Kun He, Yingru Li, Sucheta Soundarajan, and John~E Hopcroft.
\newblock Hidden community detection in social networks.
\newblock {\em Information Sciences}, 425:92--106, 2018.

\bibitem{holland1983stochastic}
Paul~W Holland, Kathryn~Blackmond Laskey, and Samuel Leinhardt.
\newblock Stochastic blockmodels: First steps.
\newblock {\em Social Networks}, 5(2):109--137, 1983.

\bibitem{hozhabrierdi2022coverd}
Pegah Hozhabrierdi and Sucheta Soundarajan.
\newblock Coverd: Community-based vertex defense against crawling adversaries.
\newblock In {\em International Conference on Complex Networks and Their
  Applications}, 2022.

\bibitem{graphMLP}
Yang Hu, Haoxuan You, Zhecan Wang, Zhicheng Wang, Erjin Zhou, and Yue Gao.
\newblock Graph-mlp: Node classification without message passing in graph.
\newblock {\em arXiv preprint arXiv:2106.04051}, 2021.

\bibitem{jalali2023fairness}
Zeinab~S Jalali, Qilan Chen, Shwetha~M Srikanta, Weixiang Wang, Myunghwan Kim,
  Hema Raghavan, and Sucheta Soundarajan.
\newblock Fairness of information flow in social networks.
\newblock {\em ACM Transactions on Knowledge Discovery from Data}, 17(6):1--26,
  2023.

\bibitem{jiang2022graph}
Weiwei Jiang and Jiayun Luo.
\newblock Graph neural network for traffic forecasting: A survey.
\newblock {\em Expert Systems with Applications}, page 117921, 2022.

\bibitem{jiang2019censnet}
Xiaodong Jiang, Pengsheng Ji, and Sheng Li.
\newblock Censnet: Convolution with edge-node switching in graph neural
  networks.
\newblock In {\em The International Joint Conference on Artificial
  Intelligence}, 2019.

\bibitem{kang2020inform}
Jian Kang, Jingrui He, Ross Maciejewski, and Hanghang Tong.
\newblock Inform: Individual fairness on graph mining.
\newblock In {\em International Conference on Knowledge Discovery and Data
  Mining}, 2020.

\bibitem{kang2021fair}
Jian Kang and Hanghang Tong.
\newblock Fair graph mining.
\newblock In {\em International Conference on Information and Knowledge
  Management}, 2021.

\bibitem{kipf2016semi}
Thomas~N Kipf and Max Welling.
\newblock Semi-supervised classification with graph convolutional networks.
\newblock {\em arXiv preprint arXiv:1609.02907}, 2016.

\bibitem{kleindessner2019guarantees}
Matth{\"a}us Kleindessner, Samira Samadi, Pranjal Awasthi, and Jamie
  Morgenstern.
\newblock Guarantees for spectral clustering with fairness constraints.
\newblock In {\em International Conference on Machine Learning}, 2019.

\bibitem{lancichinetti2008benchmark}
Andrea Lancichinetti, Santo Fortunato, and Filippo Radicchi.
\newblock Benchmark graphs for testing community detection algorithms.
\newblock {\em Physical Review E}, 78(4):046110, 2008.

\bibitem{li2021dyadic}
Peizhao Li, Yifei Wang, Han Zhao, Pengyu Hong, and Hongfu Liu.
\newblock On dyadic fairness: Exploring and mitigating bias in graph
  connections.
\newblock In {\em International Conference on Learning Representations}, 2021.

\bibitem{li2020deep}
Peizhao Li, Han Zhao, and Hongfu Liu.
\newblock Deep fair clustering for visual learning.
\newblock In {\em The Conference on Computer Vision and Pattern Recognition},
  2020.

\bibitem{lopes2020parallel}
Tales Lopes, Victor Str{\"o}ele, Mario Dantas, Regina Braga, and Jean-Francois
  Meh{\'a}ut.
\newblock A parallel graph partitioning approach to enhance community detection
  in social networks.
\newblock In {\em IEEE Symposium on Computers and Communications}, 2020.

\bibitem{louizos2015variational}
Christos Louizos, Kevin Swersky, Yujia Li, Max Welling, and Richard Zemel.
\newblock The variational fair autoencoder.
\newblock {\em arXiv preprint arXiv:1511.00830}, 2015.

\bibitem{masrour2020bursting}
Farzan Masrour, Tyler Wilson, Heng Yan, Pang-Ning Tan, and Abdol Esfahanian.
\newblock Bursting the filter bubble: Fairness-aware network link prediction.
\newblock In {\em Proceedings of the AAAI conference on artificial
  intelligence}, 2020.

\bibitem{highschool}
Rossana Mastrandrea, Julie Fournet, and Alain Barrat.
\newblock Contact patterns in a high school: a comparison between data
  collected using wearable sensors, contact diaries and friendship surveys.
\newblock {\em PloS one}, 10(9):e0136497, 2015.

\bibitem{newman2004finding}
Mark~EJ Newman and Michelle Girvan.
\newblock Finding and evaluating community structure in networks.
\newblock {\em Physical Review E}, 69(2):026113, 2004.

\bibitem{nickel2015review}
Maximilian Nickel, Kevin Murphy, Volker Tresp, and Evgeniy Gabrilovich.
\newblock A review of relational machine learning for knowledge graphs.
\newblock {\em Proceedings of the IEEE}, 104(1):11--33, 2015.

\bibitem{paszke2019pytorch}
Adam Paszke, Sam Gross, Francisco Massa, Adam Lerer, James Bradbury, Gregory
  Chanan, Trevor Killeen, Zeming Lin, Natalia Gimelshein, Luca Antiga, et~al.
\newblock Pytorch: An imperative style, high-performance deep learning library.
\newblock In {\em Advances in Neural Information Processing Systems}, 2019.

\bibitem{rahman2019fairwalk}
Tahleen Rahman, Bartlomiej Surma, Michael Backes, and Yang Zhang.
\newblock Fairwalk: towards fair graph embedding.
\newblock In {\em International Joint Conference on Artificial Intelligence},
  2019.

\bibitem{red2011comparing}
Veronica Red, Eric~D Kelsic, Peter~J Mucha, and Mason~A Porter.
\newblock Comparing community structure to characteristics in online collegiate
  social networks.
\newblock {\em SIAM review}, 53(3):526--543, 2011.

\bibitem{sahoo2021multiple}
Somya~Ranjan Sahoo and Brij~B Gupta.
\newblock Multiple features based approach for automatic fake news detection on
  social networks using deep learning.
\newblock {\em Applied Soft Computing}, 100:106983, 2021.

\bibitem{schmidt2020fair}
Melanie Schmidt, Chris Schwiegelshohn, and Christian Sohler.
\newblock Fair coresets and streaming algorithms for fair k-means.
\newblock In {\em Approximation and Online Algorithms: 17th International
  Workshop, WAOA 2019, Munich, Germany, September 12--13, 2019, Revised
  Selected Papers 17}, pages 232--251, 2020.

\bibitem{velickovic2017graph}
Petar Velickovic, Guillem Cucurull, Arantxa Casanova, Adriana Romero, Pietro
  Lio, Yoshua Bengio, et~al.
\newblock Graph attention networks.
\newblock {\em Stat}, 1050(20):10--48550, 2017.

\bibitem{SC}
Ulrike Von~Luxburg.
\newblock A tutorial on spectral clustering.
\newblock {\em Statistics and computing}, 17:395--416, 2007.

\bibitem{watts1998collective}
Duncan~J Watts and Steven~H Strogatz.
\newblock Collective dynamics of ‘small-world’networks.
\newblock {\em Nature}, 393(6684):440--442, 1998.

\bibitem{wu2021learning}
Le~Wu, Lei Chen, Pengyang Shao, Richang Hong, Xiting Wang, and Meng Wang.
\newblock Learning fair representations for recommendation: A graph-based
  perspective.
\newblock In {\em Proceedings of the Web Conference 2021}, 2021.

\bibitem{wu2020comprehensive}
Zonghan Wu, Shirui Pan, Fengwen Chen, Guodong Long, Chengqi Zhang, and S~Yu
  Philip.
\newblock A comprehensive survey on graph neural networks.
\newblock {\em IEEE Transactions on Neural Networks and Learning Systems},
  32(1):4--24, 2020.

\bibitem{yao2017beyond}
Sirui Yao and Bert Huang.
\newblock Beyond parity: Fairness objectives for collaborative filtering.
\newblock In {\em Advances in Neural Information Processing Systems}, 2017.

\bibitem{zhang2021graph}
Hongyan Zhang, Limei Lin, Li~Xu, and Xiaoding Wang.
\newblock Graph partition based privacy-preserving scheme in social networks.
\newblock {\em Journal of Network and Computer Applications}, 195:103214, 2021.

\bibitem{zhang2020deep}
Ziwei Zhang, Peng Cui, and Wenwu Zhu.
\newblock Deep learning on graphs: A survey.
\newblock {\em IEEE Transactions on Knowledge and Data Engineering},
  34(1):249--270, 2020.

\end{thebibliography}


\end{document}